\documentclass[12pt]{article}

\usepackage{amssymb,amsmath,amsfonts,eurosym,geometry,ulem,graphicx,caption,color,setspace,sectsty,comment,footmisc,caption,natbib,pdflscape,subfigure,array}
\usepackage{xcolor} 
\definecolor{lightblue}{RGB}{100,149,237}
\usepackage{longtable} 
\usepackage{amssymb} 
\usepackage{url}
\usepackage[table,xcdraw]{xcolor}
\usepackage{xcolor}
\usepackage{graphicx}
\usepackage{enumerate}
\usepackage{amsmath}
\usepackage{dsfont}
\usepackage{amsfonts, amssymb, float, placeins, tabularx, array, longtable, booktabs, listings} 
\usepackage{natbib}
\usepackage{threeparttable}
\usepackage{graphicx}
\usepackage{subcaption}
\usepackage{hyperref}
\hypersetup{
  colorlinks   = true,
  urlcolor     = blue,
  linkcolor    = black,
  citecolor    = blue
}
\newcolumntype{L}[1]{>{\raggedright\let\newline\\arraybackslash\hspace{0pt}}m{#1}}
\newcolumntype{C}[1]{>{\centering\let\newline\\arraybackslash\hspace{0pt}}m{#1}}
\newcolumntype{R}[1]{>{\raggedleft\let\newline\\arraybackslash\hspace{0pt}}m{#1}}

\begin{document}

\begin{titlepage}
\centering
\singlespacing
{\LARGE \textbf{Does Local Urban Governance Status Matter? Evidence from India} \par}

\vspace{0.8cm}

{\Large Saannidhya Rawat \par}

\vspace{0.2cm}


\vspace{0.5cm}

{\normalsize March 2026 \par}

\vspace{0.5cm}

\begin{abstract}
\onehalfspacing
\noindent 
We exploit quasi-random variation around the multi-threshold criteria used to classify Census Towns (CTs) and focus on settlements near the thresholds that are likely to obtain statutory recognition. Using a local fuzzy regression discontinuity design and a multi-threshold criteria, we show that meeting the CT eligibility in 2001 raises the probability of statutory recognition by 2011. Instrumenting statutory recognition with CT eligibility, we estimate the effects of ULB status on local public goods provision: government schools increase by 13.86 (primary), 7.72 (middle), and 4.89 (secondary) units, healthcare infrastructure expands by 2.53 hospitals and 3.00 family welfare centers, and financial access deepens with 4.09 cooperative banks and 2.84 agricultural credit societies. Community amenities also improve, while sports infrastructure declines by 5.71 facilities, consistent with reallocation of urban land. The corresponding reduced-form estimates are directionally consistent and indicate that crossing the CT eligibility frontier improves public goods provision. Our findings indicate that timely municipalization of emerging urban areas can expand provision of public goods.

\vspace{0.2cm}
\end{abstract}

\vfill

\begin{flushleft}
\noindent\rule{0.5\textwidth}{0.4pt}\\[4pt]
\textbf{Corresponding author:} Saannidhya Rawat, Department of Economics, Carl H. Lindner College of Business, University of Cincinnati. Email: \href{mailto:rawatsa@mail.uc.edu}{rawatsa@mail.uc.edu}.
\end{flushleft}

\setcounter{page}{0}
\thispagestyle{empty}
\end{titlepage}
\pagebreak \newpage
 
\doublespacing

\section{Introduction} \label{sec:introduction}

Urban local bodies (ULBs) are responsible for a wide range of public goods in developing countries such as schools, primary health facilities, sanitation, streets, lighting, and local regulation. Accordingly, whether and how local urban governance status matters for local economic development is a long-standing question, with implications for many rapidly urbanizing economies \citep{BrolloEtAl2013, Faguet2012, PatrickMothorpe2017}. However, a key empirical challenge is that governance structures are endogenous: places that are bigger and more developed are more likely to be recognized as urban towns and receive municipal status, while smaller or less developed settlements may not receive such recognition and therefore stagnate even further by being outside the urban system \citep{gadenne2017}. As a result, simple cross-sectional comparisons in such contexts will conflate pre-existing differences with the effects of policy. Put differently, governance status is chosen rather than randomly assigned, making causal inference difficult. To address this challenge, researchers have often relied on varying identification strategies. For instance, \cite{hae_nim_sunny_lee} finds a positive effect of urban governance on access to water facilities in India by developing a satellite-based measure of confounders such as urbanization and also controlling for population density and other demographic variables in an OLS framework with fixed effects. By contrast, \cite{mukhopadhyay_does_2017} finds no significant effect of governance status on public goods, such as water and sanitation, in India by comparing the provision of public goods across different types of urban settlements as categorized by their governance status. Meanwhile, although most papers study the impact on the extensive margin, more recently, \cite{narasimhan_polity_2024} focus on the intensive margin of public goods provision, examining how governance status may have differential effects on public goods provision depending on the size of the settlement. More broadly, other studies have also explored this question in different contexts and using various methods \citep{mukhopadhyay2015republic, mitra_city_2018}, yet the empirical evidence remains mixed.


This paper leverages a sharp institutional feature in India to address this identification problem: the Census Town (CT) classification thresholds that determine eligibility for an area to be considered urban. The Indian Census defines Census Towns as settlements meeting all three criteria: (i) population of at least 5,000 (ii) population density of at least 400 per square kilometer and (iii) at least 75\% of the male main workforce in non-agricultural activities. Crossing these multi-dimensional cutoffs sharply increases the probability that a settlement will be formally recognized by state governments as a Statutory Town (ST), i.e. granted a municipality or other ULB status, but notably, it \textit{does not} guarantee it. We exploit this quasi-random variation in recognition likelihood around the thresholds to construct a local fuzzy regression discontinuity (RD) design. In effect, meeting the Census Town criteria in 2001 serves as an instrument for obtaining statutory recognition in 2011, allowing us to isolate the effect of local urban governance status on public goods offered by the settlements.

The core idea is a fuzzy multi-threshold RD with multiple running variables. Using rich settlement-level microdata from the Socioeconomic High-resolution Rural-Urban Geographic (SHRUG) platform \citep{asher2019shrug}, we create a running variable called frontier distance using the three CT cutoffs and restrict attention to settlements near these thresholds. We first confirm a strong first-stage relationship: becoming marginally eligible as a Census Town in 2001 leads to a jump in the probability of statutory recognition in 2011. We then estimate the effect of statutory recognition, i.e., ULB status, on a broad set of development indicators measured by the 2011 Indian Census. Our outcome data are assembled by harmonizing the 2011 Village Directory (VD) and Town Directory (TD), which enumerate local infrastructure and amenities. We link these to demographic data from the 2011 Primary Census Abstract (PCA). This provides a comprehensive settlement-level panel of outcomes including educational facilities, health infrastructure, financial institutions, and other community amenities.

{\bf Preview of results.} We find that crossing the urban eligibility thresholds leads to a pronounced increase in the likelihood of statutory recognition. Meeting all three Census Town criteria in 2001 raises the probability of obtaining statutory recognition by 7.1 percentage points in our preferred local specification, with a first-stage F-statistic of 18.05. Using CT eligibility as an instrument for statutory town status, we find large local average treatment effects on public goods provision: ULB status increases the number of government primary schools by 13.86 per settlement, middle schools by 7.72, and secondary schools by 4.89. We also find positive effects on health and financial infrastructure: statutory recognition increases hospitals by 2.53, family welfare centers by 3.00, cooperative banks by 4.09, and agricultural credit societies by 2.84. Community amenities also improve, with positive estimates for public libraries and reading rooms, while sports infrastructure declines by 5.71 facilities, consistent with land reallocation in formalizing settlements\footnote{Due to the small magnitude of the first-stage coefficient, the IV estimates scale reduced-form effects by a relatively large factor. The corresponding reduced-form estimates are directionally consistent and reported in the appendix.}. Results are robust to bandwidth choices, and robustness checks support the validity of the design. We find no evidence of manipulation around the thresholds and detect balance in baseline characteristics\footnote{Our final regressions control for any remaining differences in these variables.}.

Our work builds on a growing literature that examines how local institutional capacity and governance shape development outcomes in low- and middle-income countries. 
\cite{besley2010state} emphasize that investments in fiscal and legal capacity are key determinants of long-run development, and that variations in local administrative effectiveness can shape the quality of service delivery. 
Relatedly, \cite{gadenne2017} shows that greater fiscal autonomy can improve accountability and service delivery when citizens can observe how funds are used, while \cite{burgess2015value} highlights how political incentives affect the allocation of public goods across regions. We contribute to this literature by providing causal evidence on a different but complementary margin: the formal transition of villages to local urban governance status. Rather than focusing on local fiscal resources, we study how the administrative act of granting municipal status itself changes public goods provision. By leveraging the multi-threshold eligibility criteria for Census Town classification as a quasi-experiment, we isolate exogenous variation in the likelihood of statutory recognition and estimate its local average treatment effect on infrastructure outcomes. This approach provides new evidence on how the extension of local urban governance status affects provision of local public goods in urbanizing economies.

This paper contributes to the literature via two main fronts. First, it adds to the literature by quantifying the effect of local urban governance status on local public goods provision in large developing economies. Although \cite{denis_toward_2010}, \cite{pradhan_unacknowledged_2013} and \cite{NIUA2020InterStateVar} have noted that many Census Towns remain under rural governance despite meeting Census's urban criteria, our study is the first to estimate the loss in public goods provision for these areas that are functionally urban but rurally governed. 
Second, by leveraging the multi-threshold criteria for Census Town classification, our paper uses a novel instrument and highlights the potential for using multi-dimensional cutoffs in a fuzzy RD design setting. This approach can be applied to other contexts where policy eligibility is determined by multiple threshold-based criteria, allowing researchers to exploit quasi-random variation in policy exposure. 



\section{Background \& Data} \label{sec:data}
\subsection{Institutional Background} \label{sec:inst_back}

India's urban classification rests on two pillars. The first consists of settlements officially designated as urban by state governments via notification under municipal law. These include municipal corporations, municipalities, and nagar panchayats\footnote{A type of town council for smaller urban areas.}, and are each known as \emph{Statutory Towns} (STs). The second consists of \emph{Census Towns} (CTs), which are settlements that satisfy Census thresholds: total population at least 5{,}000; population density of at least 400 persons per km$^2$; and at least 75 percent of the male main workforce in non-agricultural activities. CTs are urban for statistical purposes but may continue under rural governance unless notified as STs by the state governments. Crossing the CT thresholds increases the chance of statutory recognition but \textit{does not guarantee it} as states differ in municipalization policy and timing. Some states have proactive policies to incorporate new urbanizing areas, while others delay or avoid creating new ULBs due to budgetary or political considerations.

\subsection{Local Governance in India}


The Nagarpalika Act, which was the 74th Amendment to the Indian Constitution enacted in 1992, provides the legal framework for urban local governance in India \citep{MHA_74thAmendment}. This constitutional amendment is supported by state-level municipal acts, such as \citealp{GoaMunicipalities1968, GujaratMunicipalities1963, KarnatakaMunicipalities1964, MaharashtraMunicipalities1965, TelanganaMunicipalities2019} etc., that outline the specific procedures for forming Urban Local Bodies (ULBs). The entire process, from initial identification to operational ULB, can take several years and is subject to state-specific variations in policy, political will, and administrative capacity. Some states proactively municipalize urbanizing settlements, while others delay recognition due to fiscal or political constraints, leading to the phenomenon of Census Towns—settlements that are functionally urban but remain under rural governance.

\subsubsection{What comes with ULB recognition?}

When a settlement receives statutory recognition as an Urban Local Body (ULB), it undergoes a fundamental transformation in governance structure, fiscal arrangements, and administrative responsibilities. The 74th Constitutional Amendment Act of 1992 mandates that ULBs assume control over 18 functions listed in the Twelfth Schedule, including urban planning, regulation of land use, roads and bridges, water supply, public health and sanitation, fire services, urban poverty alleviation, and provision of urban amenities \citep{MHA_74thAmendment}. Crucially, ULB recognition brings changes in three key dimensions:

\textbf{Governance and Administrative Structure:} Statutory towns transition from rural governance under gram panchayats to elected municipal councils or corporations. This shift creates a dedicated urban administrative apparatus with specialized departments for engineering, health, education, and revenue collection. ULBs are headed by elected mayors or municipal chairpersons, with ward-level representation ensuring political accountability for urban service delivery.

\textbf{Fiscal Autonomy and Revenue Sources:} ULBs gain the authority to levy and collect municipal taxes, most notably property taxes, as well as user charges for water, sanitation, and other services. They also become eligible for devolved funds from state finance commissions and centrally-sponsored urban schemes such as the Jawaharlal Nehru National Urban Renewal Mission (JNNURM) and later the Smart Cities Mission and AMRUT (Atal Mission for Rejuvenation and Urban Transformation). While many ULBs remain fiscally dependent on state transfers, statutory recognition opens access to dedicated urban funding streams unavailable to rural settlements.

\textbf{Service Delivery Mandates:} Upon municipalization, ULBs inherit responsibility for providing and maintaining urban infrastructure and services. This includes establishing and operating primary schools, health dispensaries, water supply systems, sewerage and drainage networks, street lighting, solid waste management, and fire protection services. 

However, the practical impact of ULB recognition is heterogeneous. Well-functioning municipalities in states with strong urban governance traditions may deliver substantially better services, while newly formed ULBs in resource-constrained states may struggle with capacity limitations, inadequate transfers, and weak revenue collection. 


\subsubsection{How does Census Town classification affect statutory recognition?}

Although Census Town (CT) classification and statutory recognition are distinct processes that are built on different criteria and governed by different authorities, there is a strong relationship between the two and one often inspires the other \citep{NIUA2020InterStateVar, roy2018predicting}. Meeting the CT criteria signals that a settlement has urban characteristics and may warrant municipal governance. When a settlement crosses the CT thresholds, it draws attention from state urban development authorities and policymakers, who may then consider it for notification as a statutory town. The CT designation provides an objective benchmark indicating that the settlement has reached a level of population size, density, and economic activity consistent with urban areas. 
This can prompt state governments to evaluate whether the settlement is ready for municipal governance and whether it can sustain the administrative and fiscal responsibilities of a ULB. 
In fact, the central government also monitors CTs and encourages states to municipalize them, as exhorted by \cite{mohua2016} where the Ministry of Housing and Urban Affairs (MoHUA) formally urged states to consider CTs for statutory recognition.


\subsection{Data Sources and Construction} \label{sec:data_sources}

Using SHRUG data platform\footnote{version 2.1 was available at the time of our study.}, we collect data at the village and town level\footnote{This is the most granular level at which information is available and is identified using SHRUG IDs.} for our study \citep{asher2019shrug}. We also collect state-level town directories from the Government of India's Open Government Data (OGD) platform and list of Statutory Towns (STs) from National Housing Bank's (NHB) website. We highlight the main datasets used in our analysis:

\begin{itemize}
    \item \textbf{Population Census Data (2001 \& 2011):} We draw on comprehensive census data for all Indian settlements across two decennial rounds. The Primary Census Abstract (PCA) provides basic demographic characteristics including total population, gender and age composition, Scheduled Caste and Scheduled Tribe populations, literacy rates, and detailed workforce composition. These data allow us to construct the three Census Town eligibility criteria: population size, population density, and share of male main workers in non-agricultural activities. 


    \item \textbf{Urban-Rural Classification:} We use SHRUG's official urban-rural classification identifiers to define each settlement's status for the 2001 and 2011 census rounds. 

    \item \textbf{Statutory Town Lists:} We obtain comprehensive lists of all statutorily recognized urban local bodies from the National Housing Bank and state-level census town directories from the Government of India’s Open Government Data (OGD) platform. 
\end{itemize}


Our final sample consists of over 575,000 settlements in India, which includes every inhabited village and town in the 2001 and 2011 Census. Of these, approximately 3,750 have statutory urban status (municipalities of some form) as of 2011, while the remaining ~571,000 are governed as rural villages. However, many of those rural-governed settlements are urban in character. In total, about 8,946 settlements meet the Census definition of urban (either they are statutory towns or they qualify as Census Towns). This implies that roughly 1.6\% of all Indian settlements were considered urban in 2011 by either criterion, and the rest (~98.4\%) were rural villages. These figures highlight that India’s urban population is concentrated in a relatively small number of settlements – which tend to be much larger on average than the multitude of villages.

\subsection{Running Variables and Treatment Assignment}

The key running variables for our analysis are the three Census Town eligibility criteria: population size, population density, and share of male main workers in non-agricultural activities. We construct these variables using the 2001 Census data for each settlement. Specifically, we consider the following definitions for the running variables\footnote{Density plots of these variables are presented in the Appendix.}:

\begin{itemize}
    \item \textbf{Population Size ($P$):} Total population of the settlement as recorded in the 2001 Census.
    \item \textbf{Population Density ($D$):} Total population in 2001 divided by the area of the settlement in square kilometers, computed from the area measures provided in the Village and Town Directories.
    \item \textbf{Non-Agricultural Share ($N$):} The proportion of male main workers engaged in non-agricultural activities in 2001, calculated as the number of male main workers in non-agricultural sectors divided by the total number of male main workers in the settlement.
\end{itemize}

We then define the treatment variable for statutory recognition based on whether a settlement was officially designated as a statutory town (ST) in 2011. This is a binary indicator that takes the value of 1 if the settlement was recognized as an ST in 2011, and 0 otherwise. The treatment assignment is not deterministic based on the running variables alone, as some settlements that meet the Census Town criteria may not receive statutory recognition due to state-level policy decisions, while others that do not meet the criteria may be recognized due to historical or political reasons. Therefore, we use a fuzzy regression discontinuity design that exploits the discontinuities in the probability of receiving statutory recognition at the thresholds defined by the running variables to identify the causal effect of local urban governance status on local public goods provision.



\subsection{Summary Statistics}

\subsubsection{Urban/Rural Classification vs Statutory Recognition}

We define statutory recognition using the state-level files available on Indian government's Open Government Data (OGD) platform\footnote{State-level files were collected via \href{https://data.gov.in/}{https://data.gov.in/}.}. These files report the list of statutory towns (municipalities, corporations, nagar panchayats, etc.) for each state as of 2011. We merge these files with the PCA data using state and settlement names to create an indicator for statutory recognition.

\begin{table}[ht]
    \centering
    \caption{Statutory Recognition and Urban Thresholds}
    \label{tab:urban_thresholds}
    \begin{tabular}{lcc}
    \hline
    \hline
    \multicolumn{3}{c}{\textbf{Panel A: Population $\geq$ 5,000}} \\
    \hline
    & \textbf{Pop $<$ 5,000} & \textbf{Pop $\geq$ 5,000} \\
    \hline
    Not Statutory & 485,580 & 17,509 \\
    Statutory & 159 & 3,570 \\
    \hline
    \end{tabular}

    \vspace{0.3cm}

    \begin{tabular}{lcc}
    \hline
    \multicolumn{3}{c}{\textbf{Panel B: Density $\geq$ 400 per km$^2$}} \\
    \hline
    & \textbf{Density $<$ 400} & \textbf{Density $\geq$ 400} \\
    \hline
    Not Statutory & 308,943 & 194,146 \\
    Statutory & 169 & 3,560 \\
    \hline
    \end{tabular}

    \vspace{0.3cm}

    \begin{tabular}{lcc}
    \hline
    \multicolumn{3}{c}{\textbf{Panel C: Non-Agricultural Male Workers $\geq$ 75\%}} \\
    \hline
    & \textbf{Non-Ag $<$ 75\%} & \textbf{Non-Ag $\geq$ 75\%} \\
    \hline
    Not Statutory & 467,234 & 35,855 \\
    Statutory & 1,056 & 2,673 \\
    \hline
    \end{tabular}

    \vspace{0.3cm}

    \begin{tabular}{lcc}
    \hline
    \multicolumn{3}{c}{\textbf{Panel D: All Three Thresholds Combined}} \\
    \hline
    & \textbf{Did not Meet Thresholds} & \textbf{Met Thresholds} \\
    \hline
    Not Statutory & 499,708 & 3,381 \\
    Statutory & 1,194 & 2,535 \\
    \hline
    \hline
    \end{tabular}
\end{table}

Table \ref{tab:urban_thresholds} provides a descriptive sense of this process by cross-tabulating threshold eligibility in 2001 and statutory status by 2011. Several patterns stand out. First, 17,509 settlements had populations above 5,000 yet were not statutory towns by 2011, compared to 3,570 settlements that were both above 5,000 and statutorily recognized (Panel A of Table~\ref{tab:urban_thresholds}). A similar discrepancy is seen for the density criterion (Panel B), where 194,146 settlements had density above 400 per km$^2$ in 2001 but were not statutory towns by 2011, while only 3,560 settlements met the density criterion and had statutory town status. For the non-agricultural workforce criterion (Panel C), 35,855 settlements had at least 75\% non-agricultural male workers but were not statutory towns, compared to 2,673 that met this criterion in 2001 and were statutory towns by 2011. Second, looking at the combined criteria (Panel D), a total of 5,916 settlements met all three urban criteria in 2001 (3,381 not statutory + 2,535 statutory). However, only 2,535 of these were actually recognized as statutory towns by 2011, while 3,381 settlements met the criteria but remained governed as rural. Conversely, there were 1,194 statutory towns that did not meet all the criteria; these are typically older small towns or special cases that had municipal status historically despite falling short on one of the Census metrics. This highlights that satisfying the Census definition is not sufficient for urban governance; political implementation by states is crucial. It also underscores the quasi-random aspect of our design: among settlements around the thresholds, some received statutory recognition by 2011 and others did not, not purely based on merit but partly due to state-specific policies and timing. Nevertheless, as we will show in Section \ref{sec:empirical}, crossing the thresholds greatly increases the likelihood of statutory recognition, providing the basis for our fuzzy regression discontinuity approach.


\subsubsection{Treatment Assignment and Treatment Status}

Table \ref{tab:treatment_assignment} presents the distribution of settlements by Census Town (CT) classification in 2001 and statutory town (ST) status by 2011. Panel A shows the full sample of all settlements, while Panel B restricts to settlements close to the CT thresholds. 

\begin{table}[ht]
\centering
\caption{Treatment Assignment and Status}
\label{tab:treatment_assignment}
\begin{tabular}{lccc}
\hline
\hline
\multicolumn{4}{c}{\textbf{Panel A: Full Sample}} \\
\hline
\textbf{CT in 2001} & \textbf{Statutory by 2011} & \textbf{N} & \textbf{Share (\%)} \\
\hline
No  & No  & 495,215 & 98.61 \\
No  & Yes & 1,199   & 0.24 \\
Yes & No  & 3,223   & 0.64 \\
Yes & Yes & 2,542   & 0.51 \\
\hline
\textbf{Total} & & \textbf{502,179} & \textbf{100.00} \\
\hline
\end{tabular}

\vspace{0.5cm}

\begin{tabular}{lccc}
\hline
\multicolumn{4}{c}{\textbf{Panel B: Close to Threshold Sample}} \\
\hline
\textbf{CT in 2001} & \textbf{Statutory by 2011} & \textbf{N} & \textbf{Share (\%)} \\
\hline
No  & No  & 36,726 & 98.86 \\
No  & Yes & 100    & 0.27 \\
Yes & No  & 297    & 0.80 \\
Yes & Yes & 28     & 0.08 \\
\hline
\textbf{Total} & & \textbf{37,151} & \textbf{100.00} \\
\hline
\hline
\end{tabular}
\caption*{\footnotesize \textit{Notes:} ``CT in 2001'' indicates whether a settlement met all three Census Town criteria (population $\geq$ 5,000, density $\geq$ 400 per km$^2$, non-agricultural workers $\geq$ 75\%) in 2001. ``Statutory by 2011'' indicates whether the settlement had statutory town status in 2011. Panel B restricts to settlements within the bandwidth around the CT thresholds.}
\end{table}

Several patterns emerge from Table \ref{tab:treatment_assignment}. In the full sample (Panel A), the vast majority of settlements (98.6\%) neither met the CT criteria in 2001 nor received statutory recognition by 2011. Among the 5,765 settlements that met the CT thresholds in 2001, only 2,542 (44\%) gained statutory status by 2011, while 3,223 (56\%) remained under rural governance despite being functionally urban. This substantial gap between eligibility and recognition illustrates the discretionary nature of state municipalization policy. Conversely, 1,199 settlements that did not meet the CT criteria nevertheless received statutory recognition, typically reflecting historical municipal status or special circumstances. 

The close-to-threshold sample (Panel B) reveals similar patterns on a smaller scale, with 297 settlements meeting CT criteria but not receiving statutory status, and 100 settlements gaining statutory status despite not meeting the criteria. This variation around the thresholds provides the identifying variation for our fuzzy regression discontinuity design, where crossing the CT thresholds substantially increases but does not guarantee statutory recognition.

\section{Empirical Strategy} \label{sec:empirical}
In this section, we lay out the identification strategy. The goal is to estimate the causal effect of statutory recognition and ULB status on settlement outcomes.

\subsection{Quasi-random variation from Census Towns} \label{sec:emp_strategy_ct}

In this section, we show that Census Town (CT) designation leads to a discontinuous increase in the probability of statutory recognition at the thresholds. Furthermore, becoming a CT does not directly change governance status as the decision to grant a ULB is made by the state government. So theoretically, we should not expect a direct effect of CT designation on local outcomes through the local urban governance status channel unless the settlement is also statutorily recognized. One could argue that CT may be capturing the differences in agglomeration economies or other unobserved characteristics that may be correlated with local outcomes. However, close to the thresholds, these characteristics should be similar on either side of the cutoff. For a country of 1.5 billion people, a settlement with a population of 4,900 and 5,100, or density of 390 and 410, is not likely to be systematically different in terms of agglomeration economies. Furthermore, even though we cannot rule out differences in unobserved characteristics, we can test for balance in observed characteristics (e.g., literacy rate, caste composition) and run placebo tests at pseudo-thresholds to check if there are any discontinuities in outcomes at these points. First, we show that there is a strong first-stage relationship between meeting the CT thresholds and statutory recognition. We then use this quasi-random variation to instrument for statutory recognition and identify the effect of statutory recognition on local outcomes.

Below we describe how we implement this empirical strategy in practice. Let $P_i$, $D_i$, and $N_i$ denote the population, density (per km²), and non-agricultural employment share of settlement $i$. Define three binary indicators for meeting the CT thresholds:

\begin{align*}
    Z_{P,i} &= \mathds{1}\!\left\{\frac{P_i}{5000} - 1 \ge 0\right\} \\
    Z_{D,i} &= \mathds{1}\!\left\{\frac{D_i}{400} - 1 \ge 0\right\} \\
    Z_{N,i} &= \mathds{1}\!\left\{\frac{N_i}{0.75} - 1 \ge 0\right\}
\end{align*}

We define CT eligibility as $Z_i = Z_{P,i} \times Z_{D,i} \times Z_{N,i}$, which equals 1 if settlement $i$ meets all three thresholds and 0 otherwise, which we use as an instrument for statutory recognition. The idea is that settlements that meet the CT thresholds are more likely to receive statutory recognition, but the decision to grant statutory recognition is not deterministic based on the CT thresholds alone. Therefore, we can use the variation in CT eligibility to identify the causal effect of statutory recognition on local outcomes.  

Next, let $ST_i$ be an indicator for statutory recognition by 2011 i.e. whether settlement $i$ gets an Urban Local Body (ULB) or not, by 2011. Then, our object of interest is the probability of statutory recognition $P(ST_i = 1 | Z_i, X_i)$, where $X_i$ is a vector of controls. Our data enables us to estimate the following first stage regression:

\begin{equation}
ST_i = \pi_0 + \pi_1 Z_i + X_i'\beta + \delta_{s(i)} + \varepsilon_i
\label{eq:first_stage_eq}
\end{equation}

where $Z_i$ is CT eligibility in 2001 (instrument), $X_i$ is a vector of smooth functions of the running variables and other controls (e.g. $Z_{P,i}, Z_{D,i}, Z_{N,i}$, literacy, caste shares), and $\delta_{s(i)}$ are district-level fixed effects. 

\begin{figure}[H]
    \centering
    \begin{minipage}{0.48\textwidth}
        \centering
        \includegraphics[width=\textwidth]{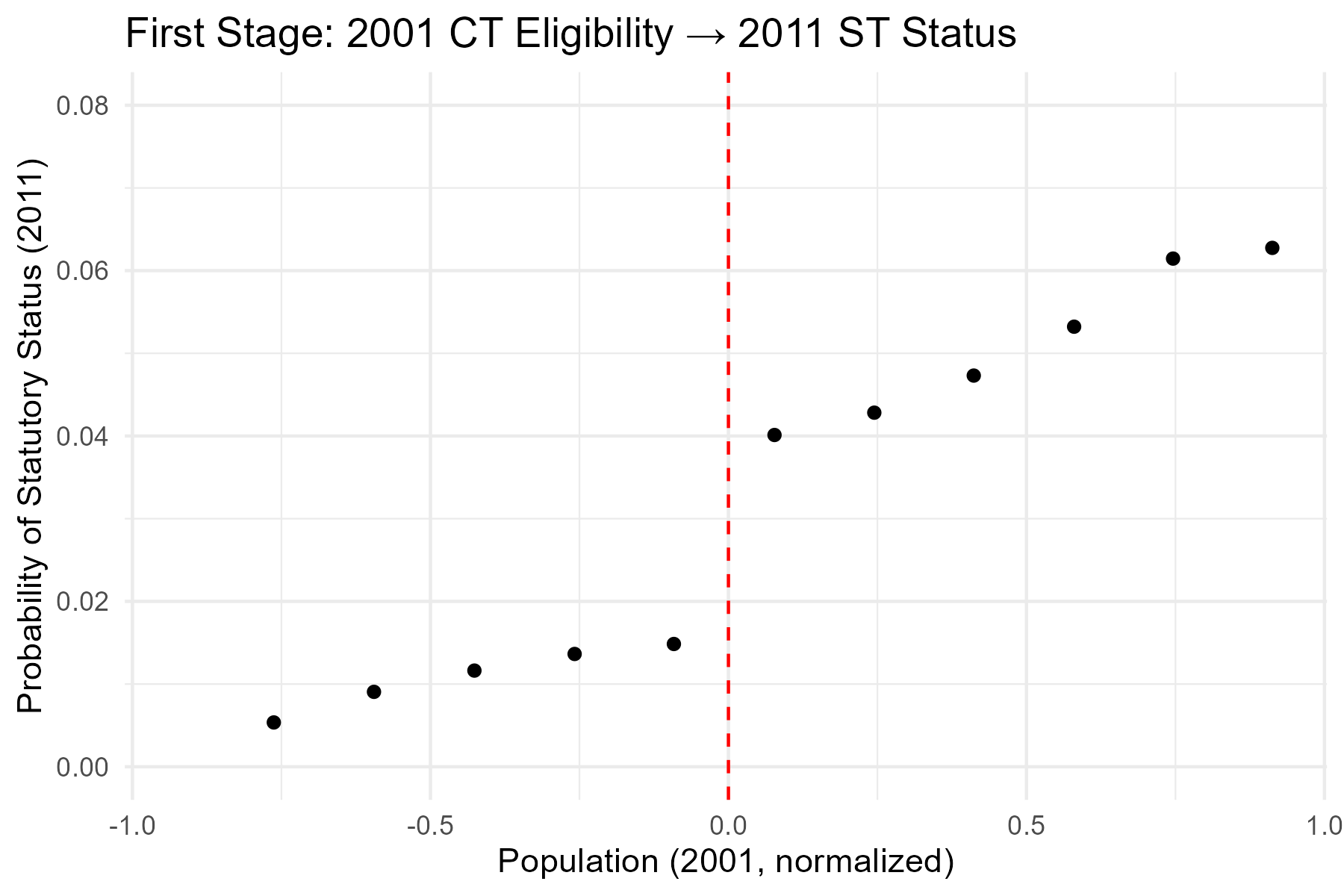}
        \caption*{(A) Population threshold}
    \end{minipage} 
    \hspace{0.5in}
    \begin{minipage}{0.48\textwidth}
        \centering
        \includegraphics[width=\textwidth]{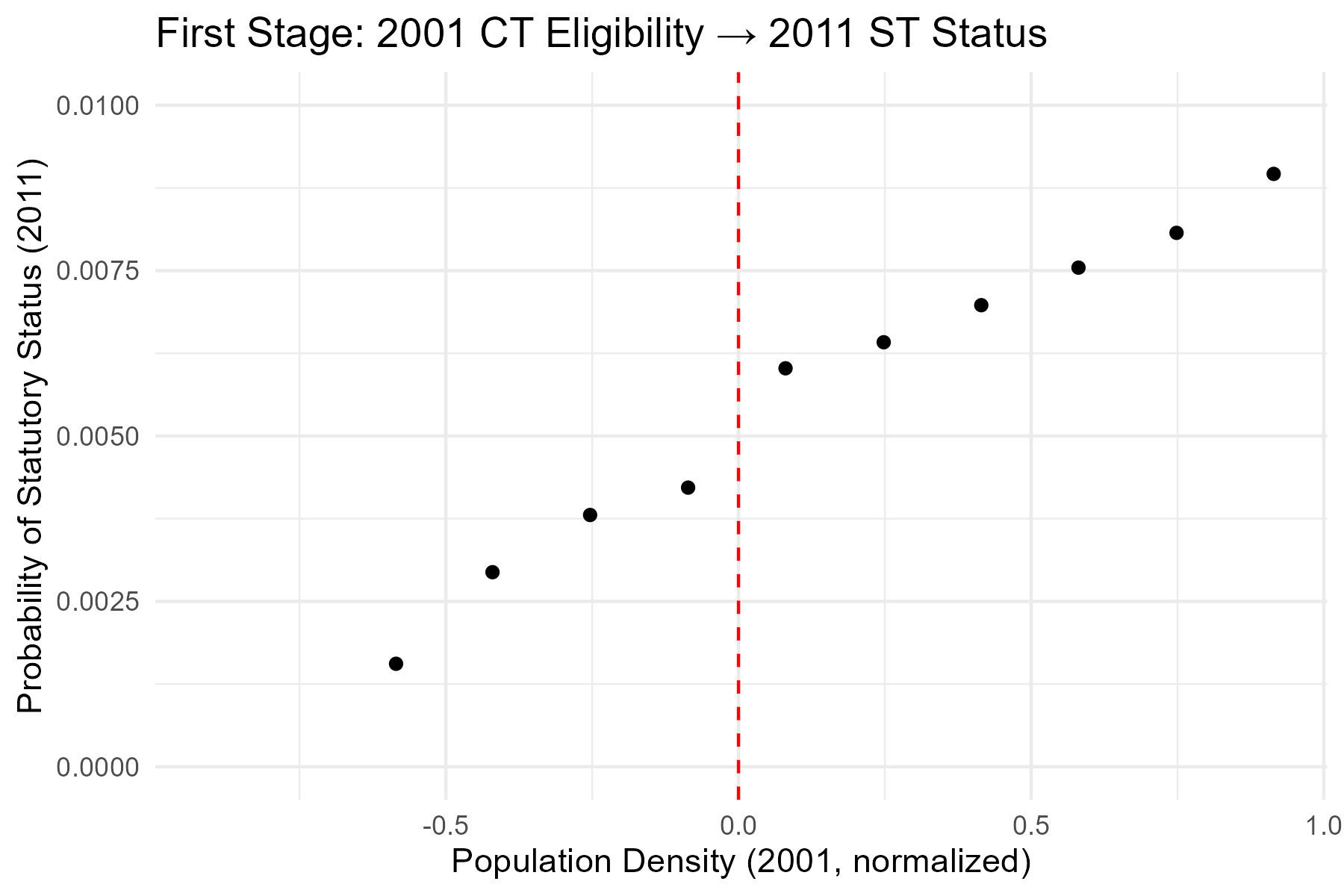}
        \caption*{(B) Density threshold}
    \end{minipage}
    
    \vspace{1em}

    \begin{minipage}{0.48\textwidth}
        \centering
        \includegraphics[width=\textwidth]{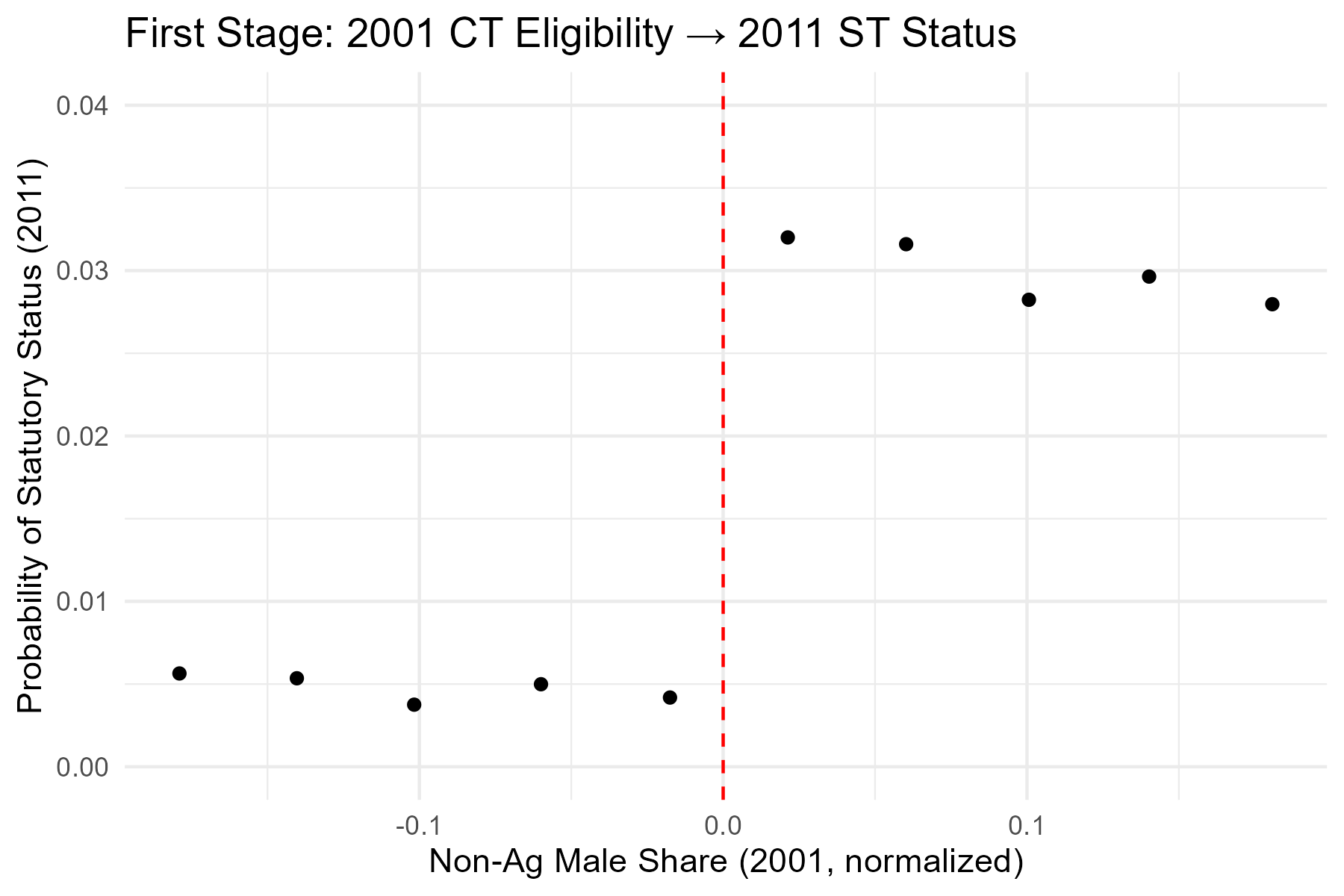}
        \caption*{(C) Non-agricultural share threshold}
    \end{minipage}
    
    \caption{First-stage Discontinuities and Statutory Recognition}
    \label{fig:first_stage}

    \medskip
    {\footnotesize \textit{Notes:} Each panel shows a binned scatter plot of the predicted probability of statutory recognition in 2011 against one of the three normalized running variables from 2001. Predicted probabilities are fitted values from the global first-stage OLS regression of statutory status on Census Town eligibility, controls, and district fixed effects (Equation \ref{eq:first_stage_eq}). The dashed red line marks the normalized threshold.}
\end{figure}

Figure \ref{fig:first_stage} presents graphically the first stage of our fuzzy RD design explained in equation \ref{eq:first_stage_eq} close to the thresholds\footnote{We present results up to 100\% away from the thresholds.}. The figure shows binned scatter plots of the probability of statutory status in 2011 against each running variable in 2001. Sub-figure (A) of Figure \ref{fig:first_stage} shows the relationship between population and statutory recognition. Probability of statutory recognition is positively related to population, which is expected since an increase in population signifies a larger settlement that is more likely to be recognized as a town. Despite this positive relationship, there is a clear discontinuity in the probability of statutory recognition at the population threshold of 0 after normalization. This discontinuity indicates that settlements that just meet the population threshold are significantly more likely to receive statutory recognition than those that just miss it. Sub-figure (B) of Figure \ref{fig:first_stage} shows the relationship between density and statutory recognition. Similar to population, there is a positive relationship between density and statutory recognition, but the discontinuity at the density threshold, while present, is less pronounced than for population. Sub-figure (C) of Figure \ref{fig:first_stage} shows the relationship between non-agricultural share and statutory recognition. The relationship is flat, but there is a clear discontinuity at the non-agricultural share threshold. Settlements that just meet the non-agricultural share threshold are more likely to receive statutory recognition than those that just miss it. Overall, these figures provide evidence of a strong first-stage relationship between meeting the CT thresholds and statutory recognition, which supports our identification strategy for estimating the causal effect of statutory recognition on local outcomes. In Section \ref{sec:results_stat}, we present a table that reports the first-stage coefficient and F-statistics for global and local estimates.

\subsection{Fuzzy RD with multiple running variables} \label{sec:emp_strategy_rd}

India’s CT thresholds create jointly-binding rules on three dimensions: population $P\ge 5{,}000$, density $D\ge 400$ per km², and male main non-agricultural share $N\ge 0.75$. Define running variables as normalized distances to the cutoffs: $r_P = P/5000 - 1$, $r_D = D/400 - 1$, and $r_N = N/0.75 - 1$. Following the logic of multi-cutoff RD, we focus on observations near the cutoffs and develop a frontier using a soft-min functional specification.

\begin{figure}[H]
	\centering
	\includegraphics[width=0.8\textwidth]{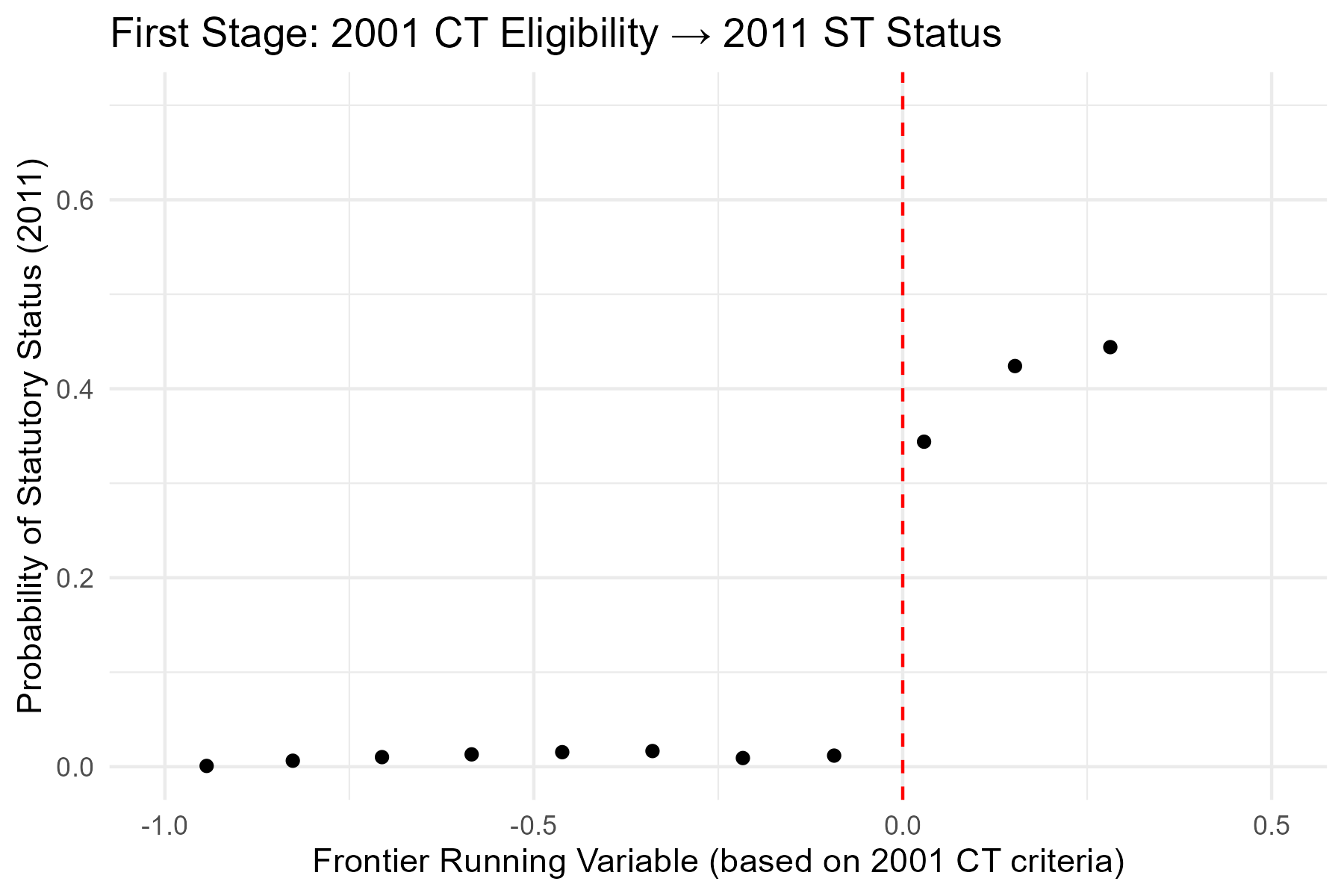}
	\caption{Probability of Statutory Recognition by Frontier Distance}
	\label{fig:pt_plot}

	\medskip
	{\footnotesize \textit{Notes:} Binned scatter plot of the predicted probability of statutory recognition in 2011 against the soft-min frontier running variable constructed from the three 2001 Census Town criteria. The frontier variable is defined as $r^*_i = -\frac{1}{\alpha}\ln\bigl(\exp(-\alpha\, r_P) + \exp(-\alpha\, r_D) + \exp(-\alpha\, r_N)\bigr)$ with $\alpha = 10$, which smoothly approximates the minimum of the three normalized distances. Negative values indicate that a settlement falls short of at least one threshold; positive values indicate that all three thresholds are met. Each dot represents the mean predicted probability within an equal-width bin. Predicted probabilities are fitted values from the global first-stage regression\footnote{For our main analysis in Section \ref{sec:results_iv}, we use local first-stage estimates.} (Equation \ref{eq:first_stage_eq}).}
\end{figure}

Figure \ref{fig:pt_plot} plots the predicted probability of statutory recognition against the frontier running variable, which collapses the three Census Town dimensions into a single scalar. The figure reveals a clear discontinuity at the eligibility frontier, near the zero on the horizontal axis. To the left of the cutoff, where settlements fall short of at least one threshold, the predicted probability of receiving a municipal body is close to zero and hovers near 1\%. Immediately to the right, where settlements satisfy all three criteria, the probability jumps sharply to approximately 35\%, rising further to above 40\% for settlements that comfortably exceed the thresholds.
This pattern is precisely what the fuzzy RD design requires: a large, discrete change in the probability of treatment at the eligibility boundary, embedded within an otherwise smooth relationship. The magnitude of the jump is substantial and visually unambiguous, providing strong graphical evidence that Census Town eligibility is a predictor of statutory recognition. This reinforces the relevance condition of our instrument and is consistent with the formal first-stage F-statistics reported in Table \ref{tab:firststage_global_local}.

The contrast between Figure \ref{fig:pt_plot} and the individual-threshold plots in Figure \ref{fig:first_stage} is also informative. When projected onto any single running variable, the discontinuity appears modest because most of the variation in eligibility is driven by the joint interaction of all three criteria. The frontier variable captures this joint variation and motivates our use of a multi-dimensional instrument defined by the intersection of all three thresholds, rather than relying on any single cutoff.

\section{Results} \label{sec:results}
\subsection{First stage: Effects on Statutory Recognition} \label{sec:results_stat}

\begin{table}[!ht]
    \centering
    \caption{First-stage: Effect of Census-Town Eligibility on Statutory Status}
    \label{tab:firststage_global_local}
    \begin{threeparttable}
    \setlength{\tabcolsep}{10pt}
    \begin{tabular}{lcc}
    \toprule
    & \textbf{Global} & \textbf{Local} \\
    \midrule
    \multicolumn{3}{l}{\textit{Dependent variable:} Statutory town $=1$}\\[3pt]

    CT eligibility in 2001 & 0.431$^{***}$ & 0.071$^{***}$ \\
    & (0.0157) & (0.0167) \\

    \addlinespace[4pt]
    Controls & Yes & Yes \\
    District fixed effects & Yes & Yes \\
    SEs clustered at district level & Yes & Yes \\

    \addlinespace[4pt]
    First-stage $F$-statistic & 749.43 & 18.05 \\
    Partial $R^2$ of instrument & 0.2522 & 0.0127 \\
    Adj. $R^2$ & 0.3493 & 0.1007 \\
    Observations & 502{,}179 & 37{,}151 \\
    Districts (FE groups) & 619 & 558 \\
    \midrule
    \multicolumn{3}{p{0.9\linewidth}}{\footnotesize \textit{Notes:} First-stage estimates of statutory recognition on the census-threshold instrument with controls and district fixed effects. Controls include population, density, non-agricultural male workforce share, literacy rate, and caste composition. Standard errors are clustered by district (in parentheses). ``Local'' restricts to settlements near the 2001 census thresholds (population $\approx$ 5{,}000, density $\approx$ 400, non-agricultural male workforce share $\approx$ 0.75). The single-IV first-stage $F$ is $t^2$ from the clustered specification.
    }\\
    \addlinespace[2pt]
    \multicolumn{3}{l}{\footnotesize $^{***}p<0.01$, $^{**}p<0.05$, $^{*}p<0.1$.}
    \\
    \bottomrule
    \end{tabular}
    \end{threeparttable}
\end{table}

The first-stage results in Table \ref{tab:firststage_global_local} reveal important patterns regarding the strength and validity of our instrumental variable strategy. In the global sample, the instrument exhibits substantial explanatory power with a coefficient of 0.431 and a first-stage F-statistic of 749.43, well exceeding conventional thresholds for weak instrument concerns. However, this strong statistical relationship may be misleading for causal identification purposes. The global specification, while statistically powerful, is potentially vulnerable to endogeneity. Settlements across the entire population and density spectrum may differ systematically in ways that correlate with both census threshold compliance and subsequent outcomes. For instance, larger settlements may have greater political influence or administrative capacity that simultaneously makes them more likely to meet census criteria and more likely to attract infrastructure investment, regardless of statutory status. 
  
In contrast, the local specification of restricting analysis to settlements near the census cutoffs\footnote{We use the following restrictions around the thresholds: (i) population $\pm 5000$ (ii) density $\pm 400$ (iii) non-ag male main work share $\pm 0.2$. Table \ref{tab:bw_sensitivity} in the appendix presents bandwidth-sensitivity results.} provides more credible identification despite the smaller coefficient magnitude (0.071) and relatively lower F-statistic (18.05). This specification focuses on settlements that are observationally similar except for small differences in census characteristics that push them just above or below the thresholds. The lower adjusted $R^2$ (0.1007) and partial $R^2$ of instrument (0.0127) indicate that we are capturing variation that is more plausibly exogenous, as settlements just above and below the cutoffs should be similar in unobservable characteristics in the absence of treatment, as long as no manipulation occurs.

The F-statistic of 18.05 for our local specification is above the conventional threshold of 16.38 for strong instruments in the just-identified case \citep{stock1997testing} and well above the critical value of 8.96 for 15\% maximal IV bias \citep{stock2002testing}, suggesting the instrument maintains reasonable strength for our identification strategy. The local estimates thus provide our preferred specification for downstream analysis, offering a credible foundation for interpreting the causal effects of statutory recognition on settlement outcomes. 

\subsection{Main effects on outcomes} \label{sec:results_iv}

We next present 2SLS estimates of the effect of statutory recognition on settlement outcomes. We group outcomes into categories most relevant to local public goods and services: education, health, financial access, and community infrastructure. The reduced-form estimates provide the numerator of the Wald ratio underlying our IV specification and show that crossing the CT eligibility frontier improves public goods provision across the same outcome categories\footnote{The full reduced-form table is reported in Appendix Table~\ref{tab:appendix_reduced_form}.}. 


\begin{table}[htbp]\centering
\caption{Effect of Local Urban Governance Status on School Provision}
\label{tab:ulb_schools}
\begin{threeparttable}
\setlength{\tabcolsep}{10pt}
\begin{tabular}{lccc}
\toprule
 & \multicolumn{3}{c}{Dependent variable: Number of government schools} \\
\cmidrule(lr){2-4}
 & Primary & Middle & Secondary \\
\midrule
Effect of ULB status & 13.86$^{***}$ & 7.72$^{***}$ & 4.89$^{***}$ \\
 & (4.00) & (2.25) & (1.30) \\
Controls & Yes & Yes & Yes \\
District FE & Yes & Yes & Yes \\
Observations & 37{,}143 & 37{,}099 & 37{,}148 \\
\bottomrule
\end{tabular}
\begin{tablenotes}
\footnotesize
\item \textit{Notes:} Table reports two-stage least squares estimates where the endogenous regressor is statutory town status (ULB $=1$) and the excluded instrument is Census Town eligibility in 2001. The local sample restricts to settlements near the 2001 census thresholds (population $\pm5{,}000$; density $\pm400$; non-agricultural male share $\pm0.20$). All specifications include controls for log population, log density, non-agricultural male workforce share, literacy rate, main worker share, and caste composition (SC and ST shares), as well as district fixed effects. Robust standard errors clustered at the district level are reported in parentheses.
\item Significance levels: $^{***}p<0.01$, $^{**}p<0.05$, $^{*}p<0.1$.
\end{tablenotes}
\end{threeparttable}
\end{table}

\paragraph{Education.} Table~\ref{tab:ulb_schools} presents the estimated effects of statutory recognition on educational infrastructure provision. The results demonstrate substantial positive impacts across all levels of schooling. Statutory recognition leads to an increase of 13.86 additional primary schools, 7.72 additional middle schools, and 4.89 additional secondary schools. All estimates are statistically significant at the 1\% level, with robust standard errors clustered at the district level. The largest absolute effect occurs at the primary level, consistent with primary education being a foundational service that local governments can expand most readily. Primary schools are also typically smaller in size and easier to open than middle or secondary schools, facilitating rapid increases in provision following statutory recognition.
As we move up the educational ladder, the estimated effects become smaller, which is plausible given the higher capital requirements, larger catchment areas, and greater regulatory demands associated with middle and secondary schools. Even so, the fact that all three levels show significant positive effects suggests that ULB status facilitates broad improvements in educational infrastructure rather than a narrow shift at only one school level.

\begin{table}[htbp]\centering
\caption{Effect of Local Urban Governance Status on Hospital Provision}
\label{tab:ulb_health}
\begin{threeparttable}
\setlength{\tabcolsep}{12pt}
\begin{tabular}{lcc}
\toprule
 & \multicolumn{2}{c}{Dependent variable} \\
\cmidrule(lr){2-3}
 & All hospitals & Family welfare centers \\
\midrule
Effect of ULB status & 2.53$^{***}$ & 3.00$^{***}$ \\
 & (0.69) & (0.88) \\
Controls & Yes & Yes \\
District FE & Yes & Yes \\
Observations & 37{,}091 & 37{,}077 \\
\bottomrule
\end{tabular}
\begin{tablenotes}
\footnotesize
\item \textit{Notes:} Two-stage least squares estimates where the endogenous regressor is statutory town status (ULB $=1$) and the excluded instrument is Census Town eligibility in 2001. The local sample restricts to settlements near the 2001 census thresholds (population $\pm5{,}000$; density $\pm400$; non-agricultural male share $\pm0.20$). All specifications include controls for log population, log density, non-agricultural male workforce share, literacy rate, main worker share, and caste composition (SC and ST shares), as well as district fixed effects. Robust standard errors clustered at the district level are reported in parentheses.
\item Significance levels: $^{***}p<0.01$, $^{**}p<0.05$, $^{*}p<0.1$.
\end{tablenotes}
\end{threeparttable}
\end{table}

\paragraph{Health.} Table~\ref{tab:ulb_health} presents the estimated effects of local urban governance status on health infrastructure provision. The results demonstrate substantial positive impacts across both health facilities. Local urban governance status leads to an increase of 2.53 additional hospitals and 3.00 additional family welfare centers. Both estimates are statistically significant at the 1\% level, with robust standard errors clustered at the district level. The positive effects on both general hospitals and family welfare centers suggest that local urban governance status facilitates improvements in core health infrastructure. The somewhat larger effect on family welfare centers is noteworthy because these facilities play a central role in maternal and child health, family planning, and preventive care within India's public health system \citep{dhingra_national_2011}. This pattern is also plausible given that family welfare centers typically require less capital and can be expanded more quickly than full-service hospitals.

\begin{table}[htbp]\centering
\caption{Effect of Local Urban Governance Status on Financial Access}
\label{tab:ulb_finance}
\begin{threeparttable}
\setlength{\tabcolsep}{10pt}
\begin{tabular}{lcc}
\toprule
 & \multicolumn{2}{c}{Dependent variable} \\
\cmidrule(lr){2-3}
 & Cooperative banks & Agricultural credit societies \\
\midrule
Effect of ULB status & 4.09$^{***}$ & 2.84$^{***}$ \\
 & (0.97) & (1.01) \\
Controls & Yes & Yes \\
District FE & Yes & Yes \\
Observations & 37{,}089 & 37{,}093 \\
\bottomrule
\end{tabular}
\begin{tablenotes}
\footnotesize
\item \textit{Notes:} Two-stage least squares estimates where the endogenous regressor is statutory town status (ULB $=1$) and the excluded instrument is Census Town eligibility in 2001. The local sample restricts to settlements near the 2001 census thresholds (population $\pm5{,}000$; density $\pm400$; non-agricultural male share $\pm0.20$). All specifications include controls for log population, log density, non-agricultural male workforce share, literacy rate, main worker share, and caste composition (SC and ST shares), as well as district fixed effects. Robust standard errors clustered at the district level are reported in parentheses.
\item Significance levels: $^{***}p<0.01$, $^{**}p<0.05$, $^{*}p<0.1$.
\end{tablenotes}
\end{threeparttable}
\end{table}

\paragraph{Financial Access.} Table~\ref{tab:ulb_finance} presents the estimated effects of local urban governance status on financial infrastructure provision. The results demonstrate substantial positive impacts across both types of financial institutions. Local urban governance status leads to an increase of 4.09 additional cooperative banks and 2.84 additional agricultural credit societies, both statistically significant at the 1\% level with robust standard errors clustered at the district level. The positive effects across both institution types indicate that municipalization enhances multiple forms of financial intermediation serving different community needs. Cooperative banks can support broader commercial and household financial activity, while agricultural credit societies remain relevant in peri-urban settlements where agricultural livelihoods persist alongside urbanization. The expansion of financial infrastructure may therefore reflect both greater demand in newly recognized towns and the possibility that local urban governance status signals administrative capacity and development potential to formal financial institutions \citep{burgess2005rural}.

\begin{table}[htbp]\centering
\caption{Effect of Local Urban Governance Status on Community Infrastructure}
\label{tab:ulb_community}
\begin{threeparttable}
\small
\setlength{\tabcolsep}{5pt}
\begin{tabular}{lcccc}
\toprule
 & \multicolumn{4}{c}{Dependent variable} \\
\cmidrule(lr){2-5}
 & Public libraries & Public reading rooms & Cinema halls & Sports infrastructure \\
\midrule
Effect of ULB status & 1.05$^{**}$ & 1.44$^{**}$ & 0.79 & $-$5.71$^{***}$ \\
 & (0.50) & (0.63) & (0.49) & (1.48) \\
Controls & Yes & Yes & Yes & Yes \\
District FE & Yes & Yes & Yes & Yes \\
Observations & 37{,}147 & 37{,}147 & 37{,}146 & 37{,}146 \\
\bottomrule
\end{tabular}
\begin{tablenotes}
\footnotesize
\item \textit{Notes:} Two-stage least squares estimates where the endogenous regressor is statutory town status (ULB $=1$) and the excluded instrument is Census Town eligibility in 2001. The local sample restricts to settlements near the 2001 census thresholds (population $\pm5{,}000$; density $\pm400$; non-agricultural male share $\pm0.20$). All specifications include controls for log population, log density, non-agricultural male workforce share, literacy rate, main worker share, and caste composition (SC and ST shares), as well as district fixed effects. Robust standard errors clustered at the district level are reported in parentheses.
\item Significance levels: $^{***}p<0.01$, $^{**}p<0.05$, $^{*}p<0.1$.
\end{tablenotes}
\end{threeparttable}
\end{table}

\paragraph{Community Infrastructure.} Table~\ref{tab:ulb_community} presents the estimated effects of local urban governance status on community infrastructure and access. The results show positive effects on several cultural and educational amenities. Local urban governance status leads to a statistically significant increase of 1.05 public libraries and 1.44 public reading rooms. The estimate for cinema halls is positive at 0.79 but imprecisely estimated. By contrast, sports infrastructure declines by 5.71 facilities. This negative effect is consistent with spatial constraints and land-use tradeoffs in urban development: as settlements formalize into statutory towns, open spaces previously used for sports facilities may be repurposed for schools, health facilities, banking infrastructure, roads, or residential development. Overall, these findings suggest that formal statutory recognition may enhance access to community facilities, though the transition may involve restructuring of certain types of infrastructure, particularly those requiring substantial land area.

\subsection{Robustness Checks}


\subsubsection{Density Plots and McCrary Test for Manipulation}

We conduct checks around the statutory recognition thresholds to assess the validity of our RD design. Specifically, we plot the density of the running variables and implement the \cite{mccrary2008manipulation} test for manipulation of the running variable, which examines the continuity of the density of the running variable at the cutoff.

\begin{figure}[H]
    \centering
    \includegraphics[width=0.8\textwidth]{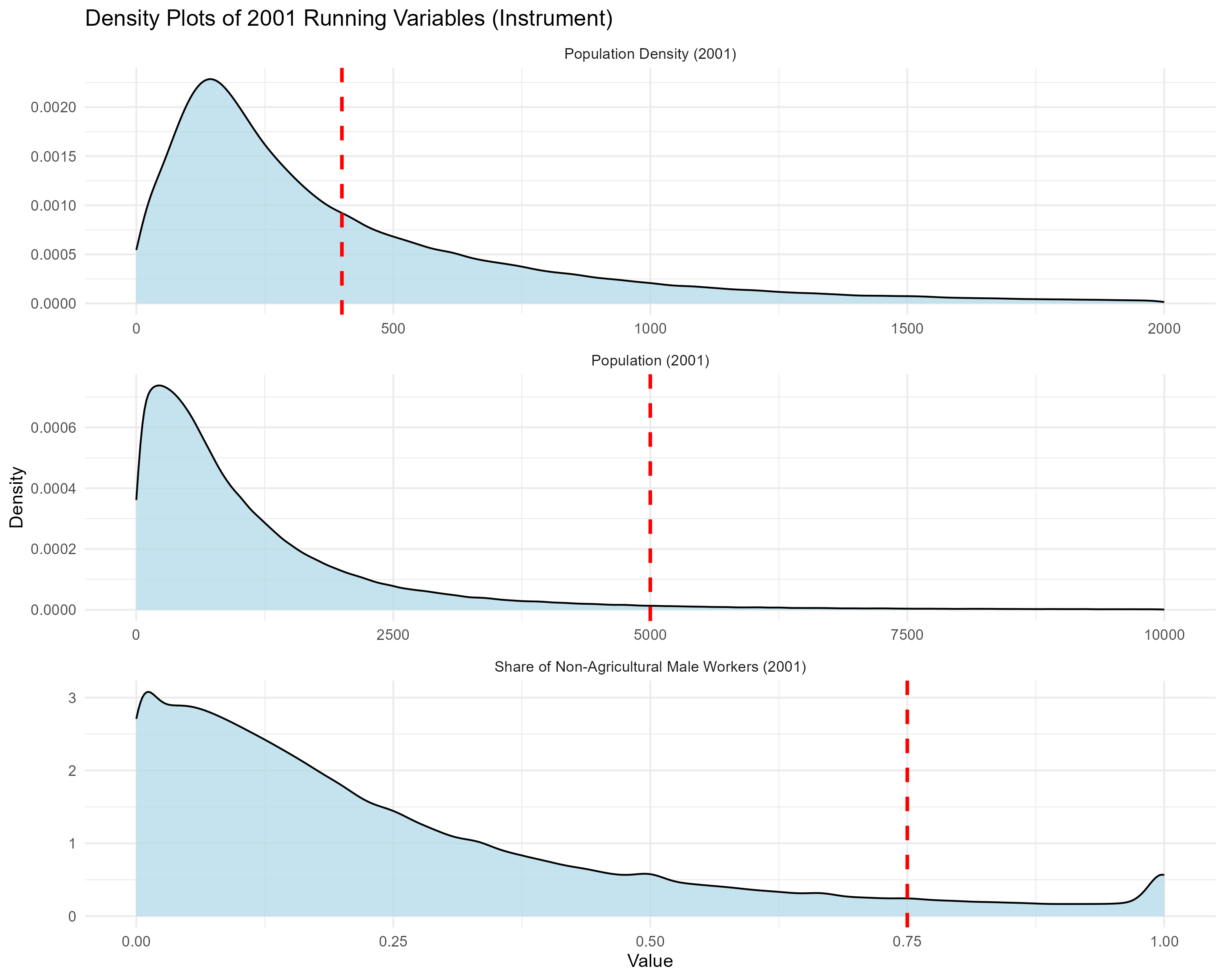}
    \caption{Density plots of CT-definition variables and thresholds}
    \label{fig:dens_plot}
\end{figure}

Figure \ref{fig:dens_plot} shows the density plots of the three running variables (population density, population, and share of non-agricultural male workers) around their respective cutoffs of 400 persons per km², 5,000 persons, and 0.75. The plots indicate that the density of the running variables is continuous and smooth around the cutoffs, with no clear signs of manipulation or strategic sorting.

\begin{table}[H]
    \centering
    \caption{McCrary Density Tests at CT Thresholds}
    \label{tab:mccrary}
    \begin{threeparttable}
    \begin{tabular}{lccc}
        \toprule
        Running Variable & Cutoff & T-statistic & P-value \\
        \midrule
        Population (2001) & 5,000 & 0.754 & 0.451 \\
        Population Density (2001) & 400 & 0.245 & 0.807 \\
        \% Non-Agricultural Male Workers (2001) & 0.75 & $-1.101$ & 0.271 \\
        \bottomrule
    \end{tabular}
    \begin{tablenotes}
        \small
        \item \textit{Notes:} This table reports results from the \cite{mccrary2008manipulation} manipulation test which examines whether there is a discontinuity in the density of the running variable at the statutory threshold. The T-statistic tests the null hypothesis of no discontinuity. All tests use a triangular kernel with bandwidth selection and jackknife variance estimation. P-values greater than 0.10 indicate no evidence of manipulation at conventional significance levels.
    \end{tablenotes}
    \end{threeparttable}
\end{table}

The \cite{mccrary2008manipulation} density test formally assesses whether there is evidence of manipulation of the running variable around the cutoff. The test estimates the density of the running variable separately on each side of the threshold and tests whether there is a discontinuity in the density at the cutoff. A significant discontinuity would suggest strategic sorting around the threshold, which would violate the identifying assumptions of the RD design.

Table \ref{tab:mccrary} presents the results of the \cite{mccrary2008manipulation} test for each of our three running variables used to define census town status. For population in 2001 (cutoff at 5,000), the test statistic is 0.754 with a p-value of 0.451. For population density in 2001 (cutoff at 400 persons per km²), the test statistic is 0.245 with a p-value of 0.807. For the share of non-agricultural male workers (cutoff at 0.75), the test statistic is $-1.101$ with a p-value of 0.271. In all three cases, we fail to reject the null hypothesis of no discontinuity in the density at conventional significance levels. This provides reassuring evidence that there is no systematic manipulation of these running variables around the statutory thresholds. Remaining robustness tests can be found in the Appendix.

\section{Conclusion} \label{sec:conclusion}

This paper provides evidence that local urban governance status leads to meaningful improvements in local public goods at India's urbanizing fringe. Using a novel multi-threshold fuzzy RD design that leverages Census Town eligibility criteria, we isolate the effect of local urban governance status on public goods provision from underlying confounding factors.

Our local IV estimates show that statutory recognition increases the number of government schools, hospitals, family welfare centers, cooperative banks, and agricultural credit societies, while decreasing the number of sports facilities, consistent with the idea of reallocation of space and land constraints in formalizing settlements. These local average treatment effects indicate that ULB status materially shifts the stock of public infrastructure in settlements near the Census Town thresholds. The corresponding reduced-form estimates are directionally consistent and confirm that crossing the eligibility frontier improves public goods provision even before scaling by the first stage.

The policy implications are twofold. First, there are clear benefits to timely municipalization of emerging urban areas. Our results show that delaying the recognition of urban settlements may imply missed opportunities for public infrastructure. Proactively converting eligible large villages into statutory towns and providing them with fiscal support and a governance framework could help improve their overall infrastructure. Second, our results highlight the importance of state capacity and support for new ULBs. We emphasize that simply declaring a settlement as a municipality is not a panacea. The improvements we document likely come from a combination of federal, state, and local initiatives, the ability to levy local taxes, greater funding availability, and improved local governance. State governments could enhance the benefits of municipalization by ensuring that newly formed ULBs receive adequate resources, technical assistance, and training to fulfill their functions.

Our study opens avenues for future research. One extension would be to examine longer-term outcomes beyond 2011, using household survey data such as subsequent rounds of
NFHS to assess impacts on education, health, and income. A clear need remains to understand \textit{why} and \textit{how} local urban governance status leads to improved public goods provision. Is it primarily through increased fiscal resources, improved administrative capacity, greater political representation, or a combination of these factors? We encourage future work to unpack the mechanisms through which municipalization translates into better infrastructure and services. Another avenue is to explore the fiscal channel: how do municipal finances change at the threshold, and to what extent do new ULBs rely on own-source revenue versus transfers? Understanding heterogeneity in statutory recognition across states, such as why some are reluctant to create new towns, possibly to avoid sharing revenue or due to political patronage networks, could inform policymakers seeking to overcome barriers to timely municipalization. Our findings underscore that formally bringing settlements into the urban administrative fold has measurable benefits, and that ensuring institutional change keeps pace with demographic change will be crucial for sustainable development in rapidly urbanizing economies.

\section*{Data and Code Availability}

{\bf Data.} All data used are public.

\noindent {\bf Code \& replication.} A complete replication package (scripts to build the running variables, construct outcomes, and reproduce figures/tables) will be submitted with the manuscript and deposited in a stable repository (e.g., Harvard Dataverse, GitHub) upon request.

\section*{Declaration of Competing Interests and Funding}

The author declares no competing interests. No external funding was received for this research.



\singlespacing
\setlength\bibsep{8pt}
\bibliographystyle{aea}

\bibliography{aej_references}

\clearpage

\appendix
\section*{Appendix} \label{sec:appendix}

\subsection*{Location of ULBs}

We map the settlements with Urban Local Bodies (ULBs) in India and present them below in Figure \ref{fig:ulb_plot}. 

\begin{figure}[H]
    \centering
    \includegraphics[width=0.8\textwidth]{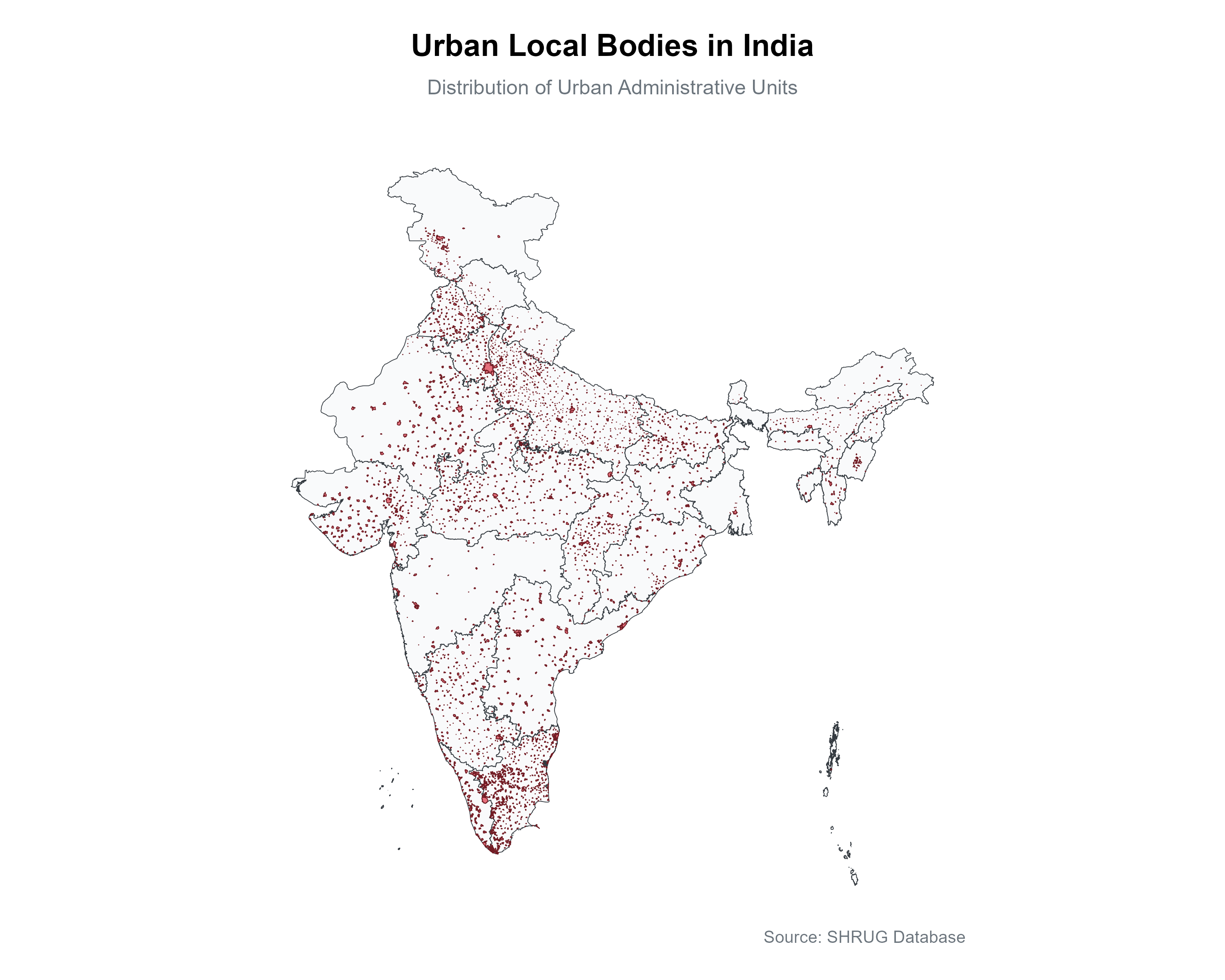}
    \caption{Urban Local Bodies (ULBs) in India}
    \label{fig:ulb_plot}
\end{figure}

\subsection*{Reduced-Form Estimates}

Table~\ref{tab:appendix_reduced_form} reports the reduced-form effect of crossing the Census Town eligibility frontier on local public goods provision in the local sample. These estimates provide the numerator of the Wald ratio underlying the main IV results in the text. The sign pattern matches the IV estimates throughout: education, health, financial access, and selected community amenities improve, while sports infrastructure declines.

\begin{table}[htbp]\centering
\caption{Reduced-Form Effects of Census Town Eligibility on Public Goods Provision}
\label{tab:appendix_reduced_form}
\begin{threeparttable}
\setlength{\tabcolsep}{8pt}
\begin{tabular}{lccr}
\toprule
Outcome & Estimate & SE & $N$ \\
\midrule
\multicolumn{4}{l}{\textit{Education}} \\
\quad Primary schools & 0.982$^{***}$ & (0.212) & 37{,}143 \\
\quad Middle schools & 0.572$^{***}$ & (0.121) & 37{,}099 \\
\quad Secondary schools & 0.346$^{***}$ & (0.066) & 37{,}148 \\
\addlinespace[4pt]
\multicolumn{4}{l}{\textit{Health}} \\
\quad All hospitals & 0.179$^{***}$ & (0.036) & 37{,}091 \\
\quad Family welfare centers & 0.212$^{***}$ & (0.041) & 37{,}077 \\
\addlinespace[4pt]
\multicolumn{4}{l}{\textit{Financial access}} \\
\quad Cooperative banks & 0.297$^{***}$ & (0.044) & 37{,}089 \\
\quad Agricultural credit societies & 0.202$^{***}$ & (0.061) & 37{,}093 \\
\addlinespace[4pt]
\multicolumn{4}{l}{\textit{Community infrastructure}} \\
\quad Public libraries & 0.074$^{**}$ & (0.030) & 37{,}147 \\
\quad Public reading rooms & 0.102$^{***}$ & (0.035) & 37{,}147 \\
\quad Cinema halls & 0.056$^{*}$ & (0.030) & 37{,}146 \\
\quad Sports infrastructure & $-$0.405$^{***}$ & (0.057) & 37{,}146 \\
\midrule
\multicolumn{4}{l}{Controls: Yes \quad District FE: Yes \quad SEs clustered by district} \\
\bottomrule
\end{tabular}
\begin{tablenotes}
\footnotesize
\item \textit{Notes:} Reduced-form estimates from OLS regressions of each outcome on Census Town eligibility in 2001 (a binary indicator for meeting all three CT thresholds). The local sample restricts to settlements near the 2001 census thresholds (population $\pm5{,}000$; density $\pm400$; non-agricultural male share $\pm0.20$). All specifications include controls for log population, log density, non-agricultural male workforce share, literacy rate, main worker share, and caste composition (SC and ST shares), as well as district fixed effects. Robust standard errors clustered at the district level are in parentheses.
\item $^{***}p<0.01$, $^{**}p<0.05$, $^{*}p<0.1$.
\end{tablenotes}
\end{threeparttable}
\end{table}

In magnitude, the reduced-form estimates are smaller because only a subset of CT-eligible settlements is ultimately statutorily recognized. Even so, the effects remain consistent with IV estimates: crossing the eligibility frontier raises the number of primary schools, middle schools, and secondary schools, while also increasing hospitals, family welfare centers, cooperative banks, and agricultural credit societies. The decline in sports infrastructure is also visible in reduced form, consistent with the land-reallocation interpretation discussed in the main text.

\subsubsection*{Balance Checks} \label{subsec:balance_checks}

To assess the validity of our regression discontinuity design, we examine whether settlements just below and just above the Census Town thresholds are comparable on observable characteristics measured in 2001. Table \ref{tab:balance_check} presents balance tests for key demographic and economic variables across both the full sample (Panel A) and the close-to-threshold sample (Panel B). The variables include log population, log density, share of non-agricultural male workers, literacy rate, main worker rate, and shares of Scheduled Caste and Scheduled Tribe populations.

\begin{table}[ht]
    \centering
    \caption{Balance Checks: 2001 variables by CT Status}
    \label{tab:balance_check}
    \small
    \begin{tabular}{@{}lccc@{}}
    \toprule
    \textbf{Variable} & \textbf{Non-CT} & \textbf{CT} & \textbf{N} \\
    \midrule
    \multicolumn{4}{l}{\textit{Panel A: Full Sample}} \\
    \addlinespace[0.2em]
    Log Population & 6.55 & 9.84 & 502,179 \\
    & (1.23) & (0.87) & \\
    Log Density & 5.69 & 7.77 & 502,179 \\
    & (1.45) & (1.12) & \\
    Non-Ag Male Workers & 0.26 & 0.90 & 502,179 \\
    & (0.24) & (0.15) & \\
    Literacy Rate & 0.47 & 0.67 & 502,179 \\
    & (0.18) & (0.12) & \\
    Main Worker Rate & 0.32 & 0.28 & 502,179 \\
    & (0.12) & (0.08) & \\
    SC Share & 0.17 & 0.13 & 502,179 \\
    & (0.21) & (0.15) & \\
    ST Share & 0.16 & 0.04 & 502,179 \\
    & (0.31) & (0.15) & \\
    \addlinespace[0.5em]
    \midrule
    \multicolumn{4}{l}{\textit{Panel B: Close-to-Threshold Sample}} \\
    \addlinespace[0.2em]
    Log Population & 6.42 & 8.85 & 37,151 \\
    & (1.18) & (0.92) & \\
    Log Density & 5.60 & 6.37 & 37,151 \\
    & (1.38) & (1.25) & \\
    Non-Ag Male Workers & 0.71 & 0.83 & 37,151 \\
    & (0.18) & (0.14) & \\
    Literacy Rate & 0.55 & 0.63 & 37,151 \\
    & (0.16) & (0.13) & \\
    Main Worker Rate & 0.27 & 0.32 & 37,151 \\
    & (0.11) & (0.09) & \\
    SC Share & 0.19 & 0.14 & 37,151 \\
    & (0.22) & (0.17) & \\
    ST Share & 0.12 & 0.05 & 37,151 \\
    & (0.27) & (0.18) & \\
    \bottomrule
    \end{tabular}
    \caption*{\footnotesize \textit{Notes:} This table presents mean characteristics in 2001 for settlements that did not meet Census Town (CT) criteria versus those that did. Standard deviations are reported in parentheses below means. Panel A shows the full sample; Panel B restricts to settlements close to the CT thresholds (within specified bandwidth). All variables are measured in 2001, prior to potential statutory recognition by 2011.}
\end{table}

As expected in the full sample (Panel A), settlements meeting CT criteria differ substantially from those that do not as they are considerably larger, denser, more economically non-agricultural, and have higher literacy rates. However, when we restrict attention to the close-to-threshold sample (Panel B), these differences narrow substantially. While some gaps remain, particularly in population and density, which are among the threshold criteria themselves, the key observation is that settlements just above and below the thresholds are much more similar than the full sample comparison would suggest. Neverthless, we control for these remaining differences in our regression models by adding these variables along with appropriate interaction terms and fixed effects.


\subsection*{Bandwidth Sensitivity}

Table~\ref{tab:bw_sensitivity} presents bandwidth sensitivity analysis for the effect on government primary schools, the outcome with the largest absolute magnitude. We vary the bandwidth multiplier from 0.5$\times$ to 2.0$\times$ the baseline thresholds and report both the reduced-form (RF) and instrumental variable (IV) estimates at each bandwidth. At narrow bandwidths (0.5$\times$, 0.75$\times$), the first stage is too weak for reliable IV estimation as we lack number of treated observations, and the reduced-form estimates are imprecise. However, starting from the baseline bandwidth (1.0$\times$) and above, the first stage becomes strong enough for valid IV estimation, and the reduced-form estimates stabilize around an increase of approximately 1 additional primary school due to crossing the CT eligibility frontier. The IV estimates at these bandwidths are also positive and statistically significant, though they vary in magnitude due to changes in the strength of the first stage and sample composition. Overall, the results suggest that our main findings are robust to reasonable variations in bandwidth choice, while very narrow bandwidths may lack sufficient power for precise estimation. 

\begin{table}[ht]
    \centering
    \caption{Bandwidth Sensitivity: Primary School Estimates}
    \label{tab:bw_sensitivity}
    \small
    \begin{tabular}{@{}lrrrcccc@{}}
    \toprule
    \textbf{Bandwidth} & $N$ & $N$(ST=1) & FS coef & RF est & RF SE & IV est & IV SE \\
    \midrule
    0.5$\times$ & 1{,}216 & 17 & 0.032 & $-$0.190 & (0.226) & $-$5.92 & (8.38) \\
    0.75$\times$ & 6{,}434 & 57 & 0.013 & 0.041 & (0.353) & 3.11 & (25.90) \\
    \textbf{1.0$\times$ (Baseline)} & \textbf{37{,}143} & \textbf{127} & \textbf{0.071} & \textbf{0.982}$^{***}$ & \textbf{(0.212)} & \textbf{13.86}$^{***}$ & \textbf{(4.00)} \\
    1.25$\times$ & 51{,}198 & 234 & 0.076 & 0.899$^{***}$ & (0.145) & 11.83$^{***}$ & (2.38) \\
    1.5$\times$ & 74{,}214 & 325 & 0.085 & 1.009$^{***}$ & (0.126) & 11.84$^{***}$ & (2.06) \\
    2.0$\times$ & 112{,}856 & 535 & 0.110 & 1.027$^{***}$ & (0.124) & 9.30$^{***}$ & (1.36) \\
    \bottomrule
    \end{tabular}
    \caption*{\footnotesize \textit{Notes:} Each row reports estimates from a different bandwidth specification. Bandwidths are multiples of the baseline thresholds: population $\pm k \times 5{,}000$, density $\pm k \times 400$, non-agricultural male share $\pm k \times 0.20$. ``FS coef'' is the first-stage coefficient on CT eligibility. ``RF'' is the reduced-form effect of CT eligibility on primary schools. ``IV'' is the 2SLS estimate of statutory recognition on primary schools. All specifications include the full set of controls and district fixed effects with district-clustered standard errors. At narrow bandwidths (0.5$\times$, 0.75$\times$), the first stage is too weak for reliable IV estimation. The reduced-form estimate is stable at approximately 1 additional primary school across all bandwidths with sufficient power ($\geq 1.0\times$). $^{***}p<0.01$.}
\end{table}

\subsection*{Descriptive Outcome Levels}

Table~\ref{tab:raw_outcome_levels} reports unadjusted outcome means by statutory status in the full sample and in the close-to-threshold local sample. These comparisons are purely descriptive and are not part of the identification argument. Their value is diagnostic: they show how large the raw full-sample differences are, and how much those differences compress once we restrict attention to settlements near the Census Town thresholds used in the main analysis.

\begin{table}[ht]
    \centering
    \caption{Descriptive Statistics: Outcome Variables by Statutory Status in the Global and Local Samples}
    \label{tab:raw_outcome_levels}
    \small
    \begin{threeparttable}
    \setlength{\tabcolsep}{4pt}
    \begin{tabular}{@{}lcccccc@{}}
    \toprule
     & \multicolumn{3}{c}{Global sample} & \multicolumn{3}{c}{Local sample} \\
    \cmidrule(lr){2-4} \cmidrule(lr){5-7}
    Outcome & Non-ST & ST & Difference & Non-ST & ST & Difference \\
    \midrule
    \multicolumn{7}{l}{\textit{Education}} \\
    \quad Primary schools & 1.27 & 15.50 & 14.23 & 1.24 & 4.96 & 3.72 \\
    \quad Middle schools & 0.55 & 7.64 & 7.09 & 0.58 & 2.65 & 2.07 \\
    \quad Secondary schools & 0.19 & 4.30 & 4.12 & 0.25 & 1.64 & 1.39 \\
    \addlinespace[4pt]
    \multicolumn{7}{l}{\textit{Health}} \\
    \quad All hospitals & 0.01 & 1.13 & 1.12 & 0.02 & 0.41 & 0.39 \\
    \quad Family welfare centers & 0.06 & 1.18 & 1.12 & 0.08 & 0.53 & 0.46 \\
    \addlinespace[4pt]
    \multicolumn{7}{l}{\textit{Financial access}} \\
    \quad Cooperative banks & 0.06 & 1.90 & 1.83 & 0.09 & 0.97 & 0.88 \\
    \quad Agricultural credit societies & 0.16 & 2.21 & 2.04 & 0.17 & 0.92 & 0.75 \\
    \addlinespace[4pt]
    \multicolumn{7}{l}{\textit{Community infrastructure}} \\
    \quad Public libraries & 0.13 & 1.02 & 0.89 & 0.11 & 0.43 & 0.31 \\
    \quad Public reading rooms & 0.15 & 1.03 & 0.88 & 0.13 & 0.45 & 0.32 \\
    \quad Cinema halls & 0.06 & 1.73 & 1.67 & 0.05 & 0.13 & 0.08 \\
    \quad Sports infrastructure & 0.42 & 1.06 & 0.64 & 0.46 & 0.49 & 0.03 \\
    \bottomrule
    \end{tabular}
    \begin{tablenotes}
    \footnotesize
    \item \textit{Notes:} Entries are unadjusted means of each outcome by statutory town status. ``Difference'' column is the raw difference in means between statutory and non-statutory settlements. The global sample uses the full regression sample with non-missing covariates and outcome data. The local sample uses the close-to-threshold sample from the main analysis. Outcome-specific sample sizes vary slightly because of missingness and match the corresponding samples used in Table~\ref{tab:ols_vs_iv}. These differences are descriptive only.
    \end{tablenotes}
    \end{threeparttable}
\end{table}

The descriptive comparison reveals a sharp contrast between the two samples. In the full sample, statutory settlements have far higher levels of nearly every public good: for example, they have 15.50 primary schools versus 1.27 in non-statutory settlements, 1.90 cooperative banks versus 0.06, and 1.73 cinema halls versus 0.06. Once we move to the local sample, these gaps narrow substantially. The compression is especially stark for community amenities: the raw gap in cinema halls falls from 1.67 to 0.08, and the raw gap in sports infrastructure shrinks from 0.64 to 0.03. Note that these raw differences are not adjusted for any covariates or fixed effects, and thus should not be interpreted as causal estimates of local urban governance. 

\subsection*{OLS versus IV Estimates}

Table~\ref{tab:ols_vs_iv} takes the comparison one step further by moving from raw descriptive differences to adjusted regression estimates. Column (1) reports global OLS estimates from the full sample of 502,110 settlements. Column (2) restricts the OLS to the close-to-threshold local sample used in the main analysis. Column (3) reports the local 2SLS estimates from Tables~\ref{tab:ulb_schools}--\ref{tab:ulb_community}, where statutory status is instrumented by Census Town eligibility. Reading alongside Table~\ref{tab:raw_outcome_levels}, the progression is straightforward: global raw differences are large, local raw differences are smaller, local OLS attenuates further after conditioning on observables, and the local IV estimates provide the paper's causal parameter.

\begin{table}[ht]
    \centering
    \caption{OLS versus IV Estimates of the Effect of Statutory Recognition}
    \label{tab:ols_vs_iv}
    \small
    \begin{threeparttable}
    \begin{tabular}{@{}lccc@{}}
    \toprule
     & (1) & (2) & (3) \\
     & OLS & OLS & 2SLS \\
     & Global & Local & Local \\
    \midrule
    \multicolumn{4}{l}{\textit{Education}} \\
    \quad Primary schools & 11.20*** & 2.08*** & 13.86*** \\
     & (0.95) & (0.49) & (4.00) \\
    \quad Middle schools & 5.22*** & 1.10*** & 7.72*** \\
     & (0.42) & (0.22) & (2.25) \\
    \quad Secondary schools & 3.14*** & 0.92*** & 4.89*** \\
     & (0.24) & (0.10) & (1.30) \\
    \addlinespace[4pt]
    \multicolumn{4}{l}{\textit{Health}} \\
    \quad All hospitals & 1.00*** & 0.36*** & 2.53*** \\
     & (0.06) & (0.06) & (0.69) \\
    \quad Family welfare centers & 0.85*** & 0.33*** & 3.00*** \\
     & (0.06) & (0.05) & (0.88) \\
    \addlinespace[4pt]
    \multicolumn{4}{l}{\textit{Financial access}} \\
    \quad Cooperative banks & 1.46*** & 0.64*** & 4.09*** \\
     & (0.10) & (0.07) & (0.97) \\
    \quad Agricultural credit societies & 1.47*** & 0.47*** & 2.84*** \\
     & (0.26) & (0.09) & (1.01) \\
    \addlinespace[4pt]
    \multicolumn{4}{l}{\textit{Community infrastructure}} \\
    \quad Public libraries & 0.51*** & $-$0.01 & 1.05** \\
     & (0.06) & (0.06) & (0.50) \\
    \quad Public reading rooms & 0.49*** & $-$0.08 & 1.44** \\
     & (0.06) & (0.07) & (0.63) \\
    \quad Cinema halls & 1.49*** & $-$0.01 & 0.79 \\
     & (0.11) & (0.04) & (0.49) \\
    \quad Sports infrastructure & $-$0.09 & $-$0.46*** & $-$5.71*** \\
     & (0.09) & (0.08) & (1.48) \\
    \midrule
    Controls & Yes & Yes & Yes \\
    District FE & Yes & Yes & Yes \\
    Observations & 502,110 & 37,148 & 37,148 \\
    \bottomrule
    \end{tabular}
    \begin{tablenotes}
    \footnotesize
    \item \textit{Notes:} Columns (1) and (2) report OLS estimates of the coefficient on statutory town status ($ST_i$). Column (3) reports 2SLS estimates where $ST_i$ is instrumented by Census Town eligibility in 2001 ($Z_i$). All specifications include controls for log population, log density, non-agricultural male workforce share, literacy rate, main worker rate, and caste shares, as well as district fixed effects. Robust standard errors clustered at the district level are in parentheses. The ``Global'' sample includes all settlements with non-missing covariates. The ``Local'' sample restricts to settlements within $\pm 5{,}000$ of the population threshold, $\pm 400$ of the density threshold, and $\pm 0.20$ of the non-agricultural share threshold.
    \item Significance: *** $p<0.01$, ** $p<0.05$, * $p<0.1$.
    \end{tablenotes}
    \end{threeparttable}
\end{table}

Taken together, Tables~\ref{tab:raw_outcome_levels} and~\ref{tab:ols_vs_iv} reveal three instructive patterns. First, both the raw means and the global OLS estimates in Column (1) show very large positive differences between statutory and non-statutory settlements for nearly all outcomes. Those differences are informative about broad urban-rural contrasts, but they are heavily contaminated by selection: larger, denser, and more economically developed settlements are both more likely to be statutory towns and more likely to have schools, hospitals, and banks, regardless of their governance status. The global comparison therefore conflates the effect of statutory recognition with pre-existing differences across places.

Second, restricting the comparison to the close-to-threshold local sample sharply compresses the differences. This is visible in the raw means above and again in the adjusted local OLS estimates in Column (2). For education outcomes, the OLS coefficient on primary schools drops from 11.20 to 2.08, and for middle schools from 5.22 to 1.10. For community infrastructure, the local OLS estimates are essentially zero --- public libraries ($-0.01$), reading rooms ($-0.08$), and cinema halls ($-0.01$) show no conditional association with statutory status. This attenuation is exactly what we would expect if settlements near the thresholds are much more comparable than the full sample.

Third, Column (3) shows that the local IV estimates that are interpreted as our paper's causal estimates for settlements whose statutory recognition status was shifted by crossing the Census Town thresholds. These 2SLS coefficients are uniformly larger than the local OLS estimates, and for several outcomes they are larger than even the global OLS estimates. For example, the IV estimate for family welfare centers (3.00) is more than 3.5 times the global OLS (0.85), and the IV estimate for public libraries (1.05) is significant where the local OLS is essentially zero. This pattern is consistent with a LATE interpretation: statutory recognition has especially large effects for marginal settlements near the eligibility frontier, while naive cross-sectional comparisons either mix together incomparable places or understate the effect because of attenuation in the observed status measure.

\clearpage





\clearpage


\end{document}